\documentclass[12pt,twoside]{article}
\usepackage{fleqn,espcrc1}

\usepackage{graphicx}

\title{Elliptic flow in Au+Au collisions at $\sqrt{s_{_{NN}}}$ = 130 GeV }

\author{R.J.M. Snellings\address{Lawrence Berkeley National Laboratory, \\
    1 Cyclotron Road, Berkeley, California 94720, United States} 
  for the STAR collaboration\footnote{For complete author list see
    J.W. Harris, these proceedings}
 }
       
\begin{document}

\maketitle

\begin{abstract}
We report the elliptic flow of charged and identified 
particles at mid-rapidity in Au+Au collisions at 
$\sqrt{s_{_{NN}}}=130$~GeV using the STAR TPC at RHIC\@. 
The integrated elliptic flow signal, $v_2$, for charged particles
reaches values of about 0.06, indicating a higher degree of
thermalization than at lower energies.
The differential elliptic flow signal, $v_2$($p_t$) up to 
1.5 GeV/$c$, shows a behavior expected from hydrodynamic model
calculations. 
Above 1.5 GeV/$c$, the data deviate from the hydro predictions;
however the $v_2$($p_t$) is still large, suggesting finite asymmetry
for the products of hard scattering.
For the identified particles, elliptic flow as a function of $p_t$ and
centrality differ significantly for particles of different masses.
This dependence can be accounted for in hydrodynamic models,
indicating that the system created shows a behavior consistent with
collective hydrodynamical flow.
\end{abstract}

\section{Introduction}
The goal of the ultra-relativistic nuclear collision program is the
creation of a system of deconfined quarks and gluons~\cite{these}. The
azimuthal anisotropy of the transverse momentum distribution for
non-central collisions is thought to be sensitive to the
early evolution of the system. The second Fourier coefficient of this
anisotropy, $v_{2}$, is called elliptic flow. It is an important
observable since it is sensitive to the rescattering of the
constituents in the created hot and dense matter. This rescattering
converts the initial spatial anisotropy, due to the almond shape of
the overlap region of non-central collisions, into momentum
anisotropy. The spatial anisotropy is largest early in the evolution of
the collision, but as the system expands and
becomes more spherical, this driving force quenches itself. Therefore,
the magnitude of the observed elliptic flow reflects the extent of the
rescattering at early time~\cite{sorge}.

Elliptic flow in ultra-relativistic nuclear collisions was first
discussed in Ref.~\cite{olli92} and has been studied intensively in
recent years at AGS~\cite{e877flow2,e895},
SPS~\cite{na49prl,na49flow,wa98} and RHIC~\cite{starflow} energies.
The studies at the top AGS energy and at SPS energies have found that 
elliptic flow at these energies is in the plane defined by the beam 
direction and 
the impact parameter, $v_2>0$, as expected from most models. 

The Solenoidal Tracker At RHIC (STAR)~\cite{STAR} measures charged
particles, and due to its azimuthal symmetry and large coverage, it is 
ideally suited for measuring elliptic flow.
The detector consists of several sub-systems in a large solenoidal
magnet. The Time
Projection Chamber (TPC) covers the pseudorapidity range $|\eta| <
1.8$ for collisions in the center of the TPC\@. The magnet is operated
at a 0.25 Tesla field, allowing tracking of particles with
$p_t>75$~MeV/c. Two Zero Degree Calorimeters~\cite{ZDC} located at
$\theta <$ 2 mrad, which mainly detect fragmentation neutrons, are
used in coincidence for the trigger. The TPC is surrounded by a
scintillator barrel which measures the charged particle multiplicity
for triggering purposes within $|\eta| < 1$. 

For this analysis, 120,000 events were selected with a primary vertex
position within 75~cm longitudinally of the TPC center and within 1~cm
radially of the beam line. For determination of the event plane,
charged particle tracks were selected with $0.1 < p_t \le 2.0$ GeV/c.
The tracks used to determine the event plane and the tracks correlated
with the event plane passed within 2~cm of the primary vertex and had
at least 15 measured space points. Also, the ratio of the number of space
points to the expected maximum number of space points for that
particular track was required to be greater than 0.52, 
suppressing split tracks from being counted more than once. 
The tracks used
for the determination of the reaction plane were within $|\eta| <
1.0$, and the tracks correlated with the reaction plane were within
$|\eta| < 1.3$.  These cuts are similar to the ones used in
Ref.~\cite{starflow} where it was shown that the analysis results are
not sensitive to the cuts.

\section{Charged-particle elliptic flow}
\label{sec:charged}

The flow analysis method~\cite{meth} involves the calculation of the
event plane angle, which is an experimental estimator of the real
reaction-plane angle.
\vspace{-0.7cm}
\begin{figure}[ht]
  \begin{minipage}[t]{0.49\textwidth}
    \includegraphics[width=1.1\textwidth]{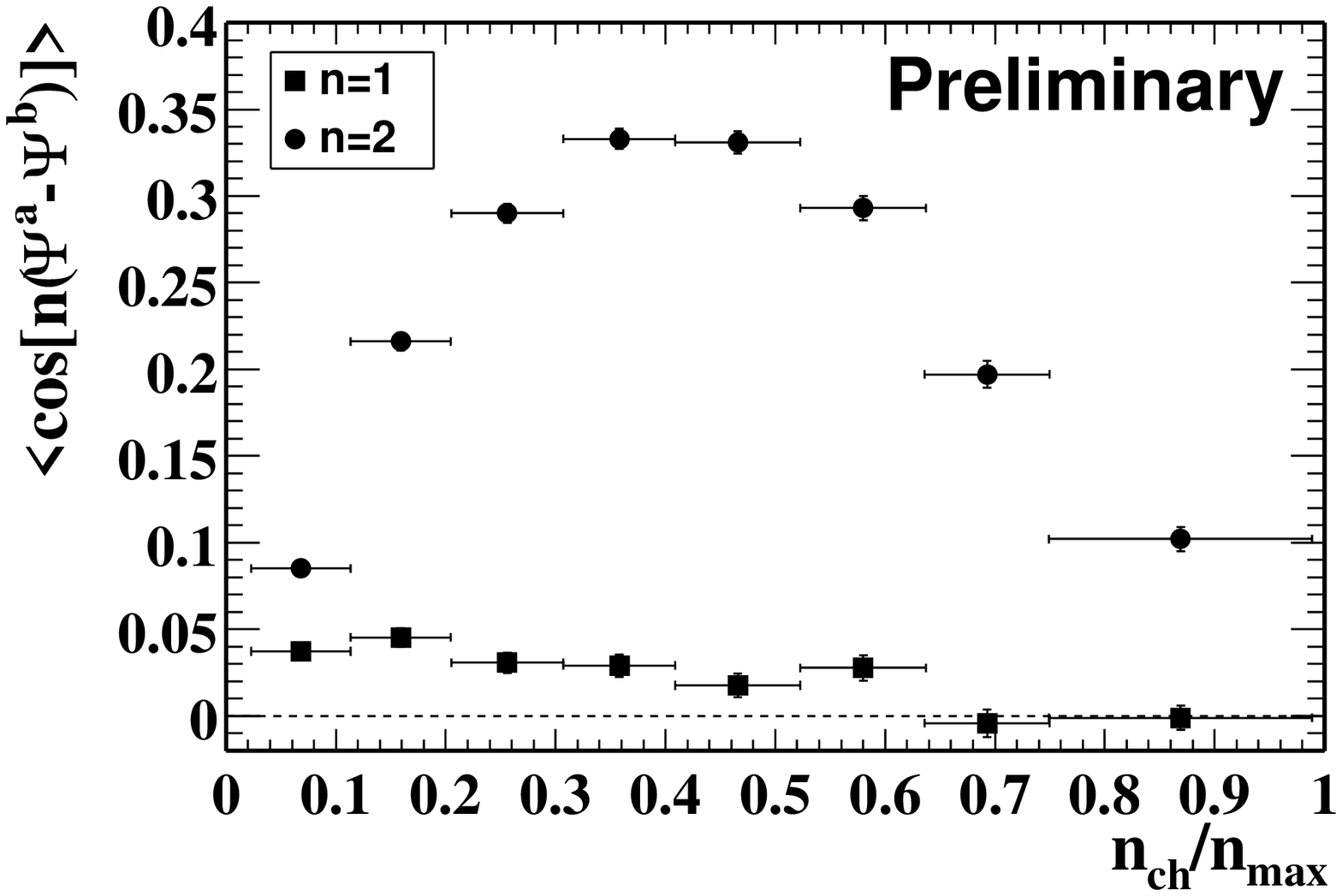}
    \caption{The correlation between the event plane 
      angles determined for
      two independent sub-events. The correlation is calculated for the
      first harmonic (n=1) and the second harmonic (n=2).}
    \label{resolution}
  \end{minipage}
  \hspace{\fill}
  \begin{minipage}[t]{0.49\textwidth}
    \includegraphics[width=1.1\textwidth]{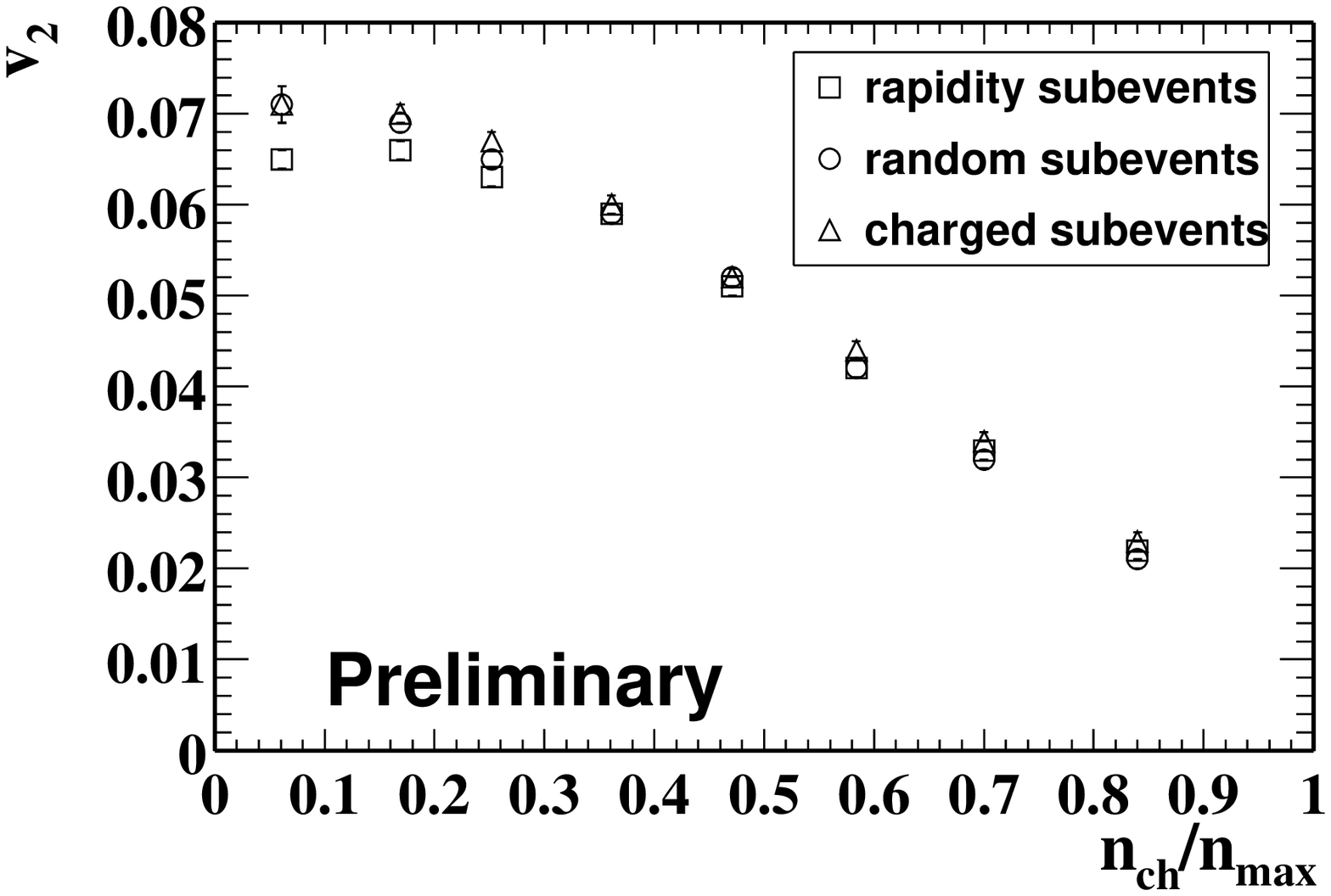}
    \caption{The integrated elliptic flow signal, $v_2$, determined using
      three different sub-event methods.}
    \label{methods}
\end{minipage}
\end{figure}
\vspace{-0.7cm}
The second harmonic event plane angle, $\Psi_2$, is calculated for
two sub-events, which are independent subsets of all tracks in each
event. 
Figure~\ref{resolution} shows the results for the correlation
between the sub-events for the first and second harmonic as a function
of centrality~\cite{starflow}. The peaked shape of the centrality
dependence of $\langle\cos[2(\Psi_a - \Psi_b)]\rangle$ is a 
signature of anisotropic flow. However, the
correlation between the sub-events may not be due entirely to
anisotropic flow. To estimate the magnitude of non-flow
effects we have chosen the sub-events in three different ways: 
1) Assigning particles with pseudorapidity $-1 < \eta < -0.05$ to one
sub-event and particles with $0.05 < \eta < 1$ to the other. Short
range correlations, such as Bose-Einstein or Coulomb, are to a large
extent eliminated by the ``gap'' between the two sub-events.
2) Dividing randomly all particles into two sub-events, sensitive to
all non-flow effects.
3) Assigning positive particles to one sub-event and negative particles
to the other, allowing an estimation of the contribution from
resonance decays. In Fig.~\ref{methods} the resulting $v_2$ versus
centrality from each of these methods is shown. The charged particles
were integrated over $0.1 < p_t < 2.0$ GeV/$c$ and $|\eta| < 1.3$. 
The results from the three methods are for the central and
mid-peripheral events very similar. However, for the most
peripheral events the results vary among the methods by about 0.005.

\vspace{-0.7cm}
\begin{figure}[ht]
  \begin{minipage}[t]{0.49\textwidth}
    \includegraphics[width=1.1\textwidth]{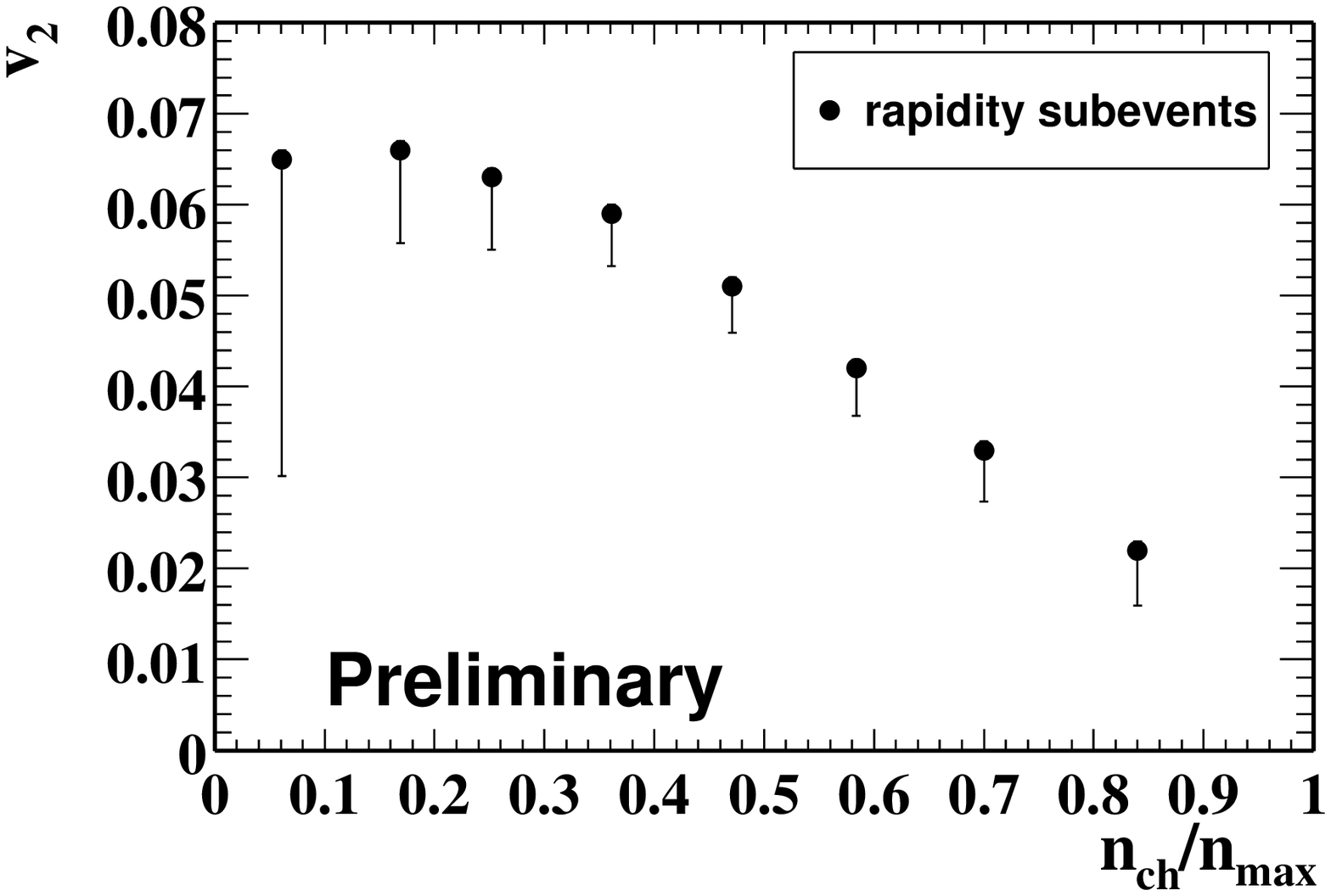}
    \caption{The integrated elliptic flow signal, $v_2$, with the
      estimated systematic uncertainties.}
    \label{errors}
  \end{minipage}
  \hspace{\fill}
  \begin{minipage}[t]{0.49\textwidth}
    \includegraphics[width=1.1\textwidth]{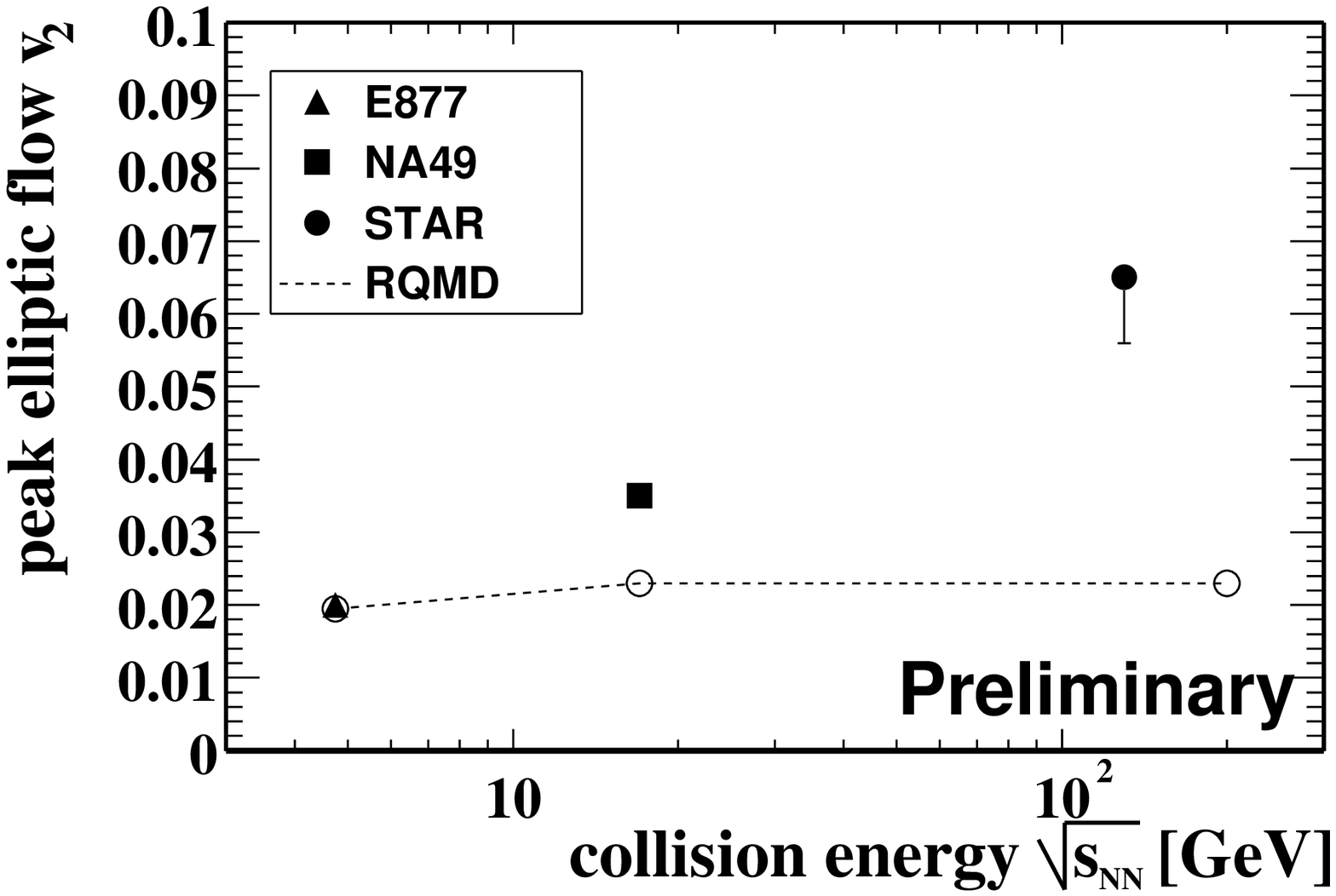}
    \caption{Excitation function of $v_2$ from top AGS to RHIC energies.}
    \label{exitation}
  \end{minipage}
\end{figure}
\vspace{-0.7cm}

Not all non-flow contributions might be known and the effects of
others, such as jets, 
are difficult to estimate because of their long-range correlation. 
Recently new methods to estimate these non-flow effects have become
available~\cite{borghini}. 
In order to estimate the systematic uncertainty due to the effects of
jets in this analysis, we assume that jets contribute at the same
level to both the first and second order correlations. This assumption
is verified by the Hijing~\cite{hijing} model.
Taking the maximum observed
positive first order correlation, 0.05, as being completely due to non-flow 
will reduce the calculated $v_2$ values.
Figure~\ref{errors} shows $v_2$ versus
centrality, where the statistical uncertainties are smaller than the markers
and the uncertainties shown are the systematic uncertainties due to
this estimated non-flow effect.

Figure~\ref{exitation} shows the maximum $v_2$ value as a function of
collision energy. It rises monotonically from about 0.02 at the top AGS
energy~\cite{e877flow2}, 0.035 at the SPS~\cite{na49flow} to about 0.06
at RHIC energies~\cite{starflow}. This increasing magnitude of the
integrated elliptic flow indicates that the degree of thermalization,
which is associated with the amount of rescattering, is higher at the
higher beam energies. However, interpretation of the excitation function
has to be done with care. The $v_2$ values used here are the maximum
values as a function of centrality for each energy. 
The centrality where $v_2$ peaks
can change as a function of beam energy, indicating different
physics~\cite{physeflow}.
 
\subsection{Differential elliptic flow}

The differential elliptic flow is a function of particle mass, $\eta$
and $p_t$. Within $|\eta| < 1.3$, $v_2$ is approximately constant. 
In this section the $v_2$($p_t$) will be discussed, the next section 
will discuss the mass dependence.

\vspace{-0.7cm}
\begin{figure}[ht]
  \begin{minipage}[t]{0.49\textwidth}
    \includegraphics[width=1.1\textwidth]{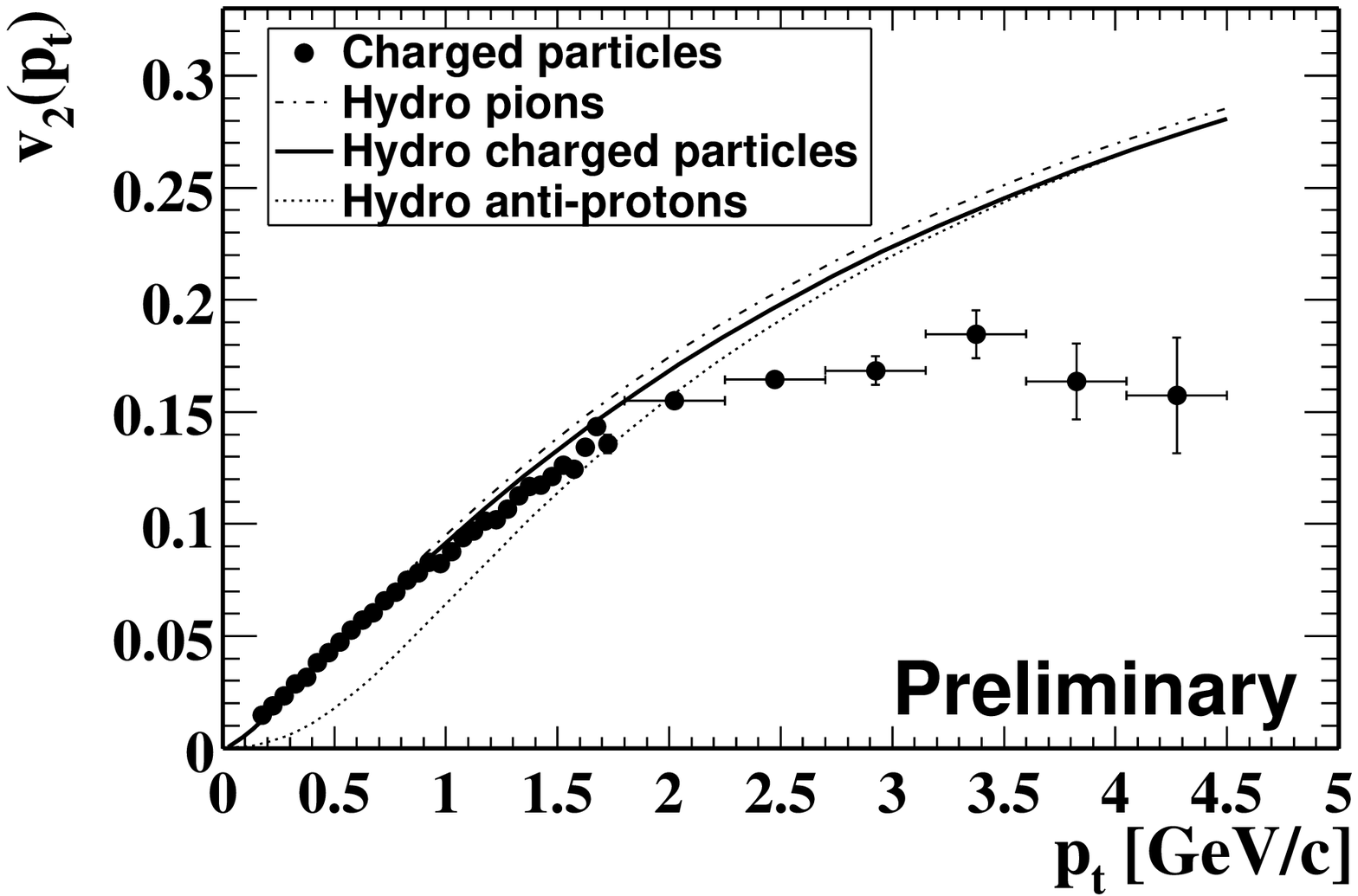}
    \caption{$v_2$($p_t$) for charged particles and minimum-bias events, 
      compared to hydro calculations~\cite{pasi2}.}
    \label{highpt_hydro}
  \end{minipage}
  \hspace{\fill}
  \begin{minipage}[t]{0.49\textwidth}
    \includegraphics[width=1.1\textwidth]{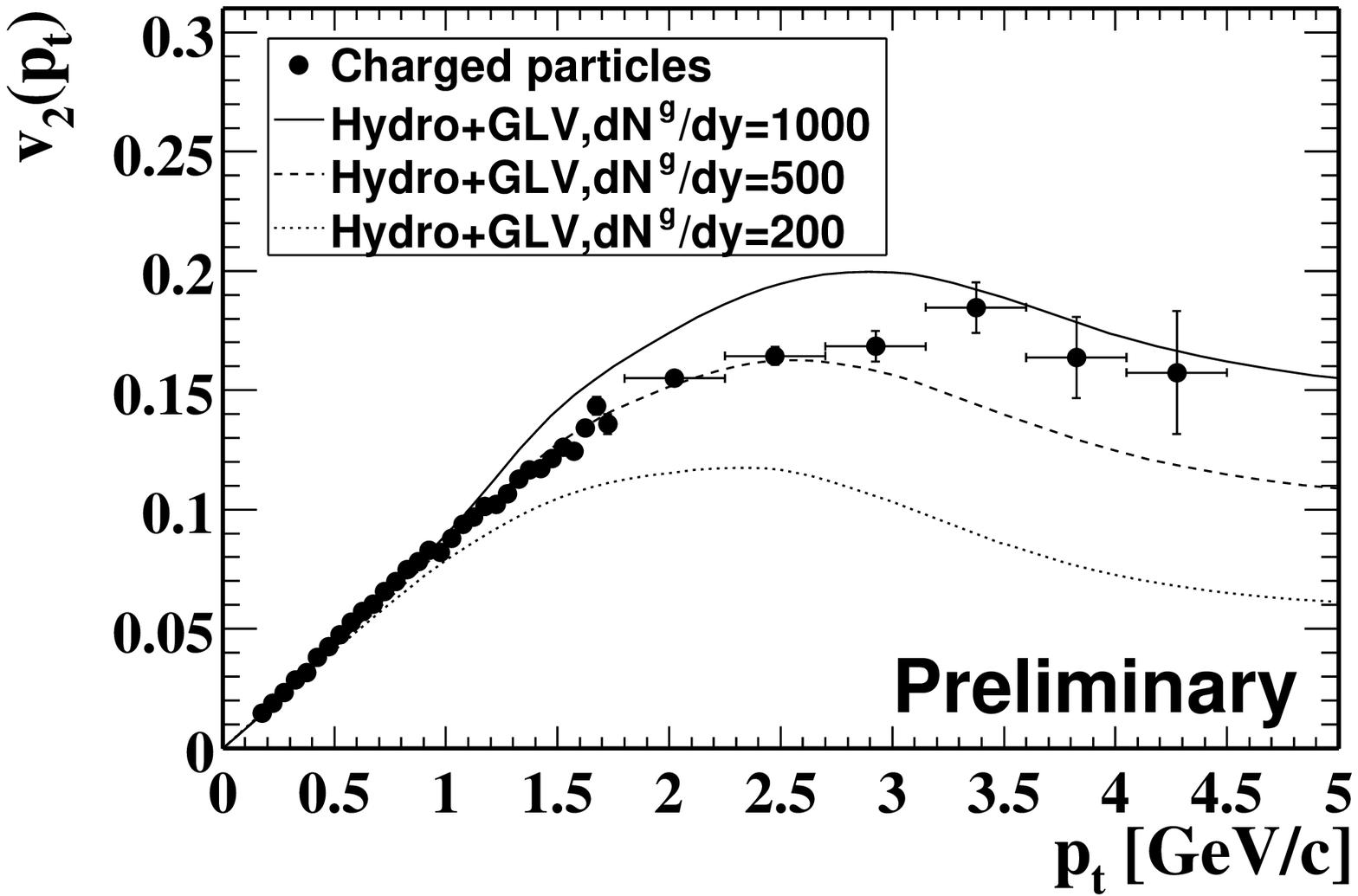}
    \caption{$v_2$($p_t$) for charged particles and minimum-bias events, 
      compared to pQCD calculations~\cite{gyulassy}.}
    \label{highpt_pqcd}
  \end{minipage}
\end{figure}
\vspace{-0.7cm}

Figure~\ref{highpt_hydro} shows $v_2$($p_t$) for charged particles,
the error bars shown are only the statistical uncertainties. 
The systematic uncertainties are 13\% up to 2 GeV/c, increasing 
to 20\% at 4.5 GeV/c.
The data show that up to $p_t < 1.5$ GeV/$c$, $v_2$($p_t$) increases
almost linearly.
This is consistent with a stronger ``in-plane'' expansion than the
average radial expansion. In Fig.~\ref{highpt_hydro}, a comparison is
made with a full hydrodynamical model calculation~\cite{pasi2}, which
describes the data up to $p_t = 1.5$ GeV/$c$. 
Above $p_t = 1.5$ GeV/$c$ the $v_2$($p_t$) starts to saturate and
deviates from the hydro calculation.
The $v_2$($p_t$) above 2 GeV/$c$ can be described by medium induced
radiative energy loss of high $p_t$ partons (jet
quenching)~\cite{gyulassy}. However, one should note that
the interplay between the ``soft'' physics and the onset of ``hard''
scattering without any energy loss could lead to a similar
behavior. In the $p_t$ range around 2--3 GeV/$c$ both ``soft'' and
``hard'' physics are expected to contribute. Future measurements at
$p_t >$ 6 GeV/$c$ should be able to disentangle these interpretations.
Figure~\ref{highpt_pqcd} shows one of the calculations~\cite{gyulassy},
which contains a phenomenological soft ``hydrodynamic'' component
combined with a hard ``pQCD'' component incorporating energy
loss. Comparing this calculation using different initial gluon densities
with the data, shows a qualitative agreement up to the highest
measured $p_t$.

\section{Identified-particle differential elliptic flow}

Hydrodynamics assumes complete local thermalization at the formation
of the system followed by an evolution governed by an Equation Of
State (EOS). 
Studies of the mass dependences of elliptic flow for particles with
$p_t <$ 1.5 GeV/$c$ provide important additional test of the
hydrodynamical model.

\begin{figure}[ht]
  \begin{minipage}[t]{0.49\textwidth}
    \includegraphics[width=1.1\textwidth]{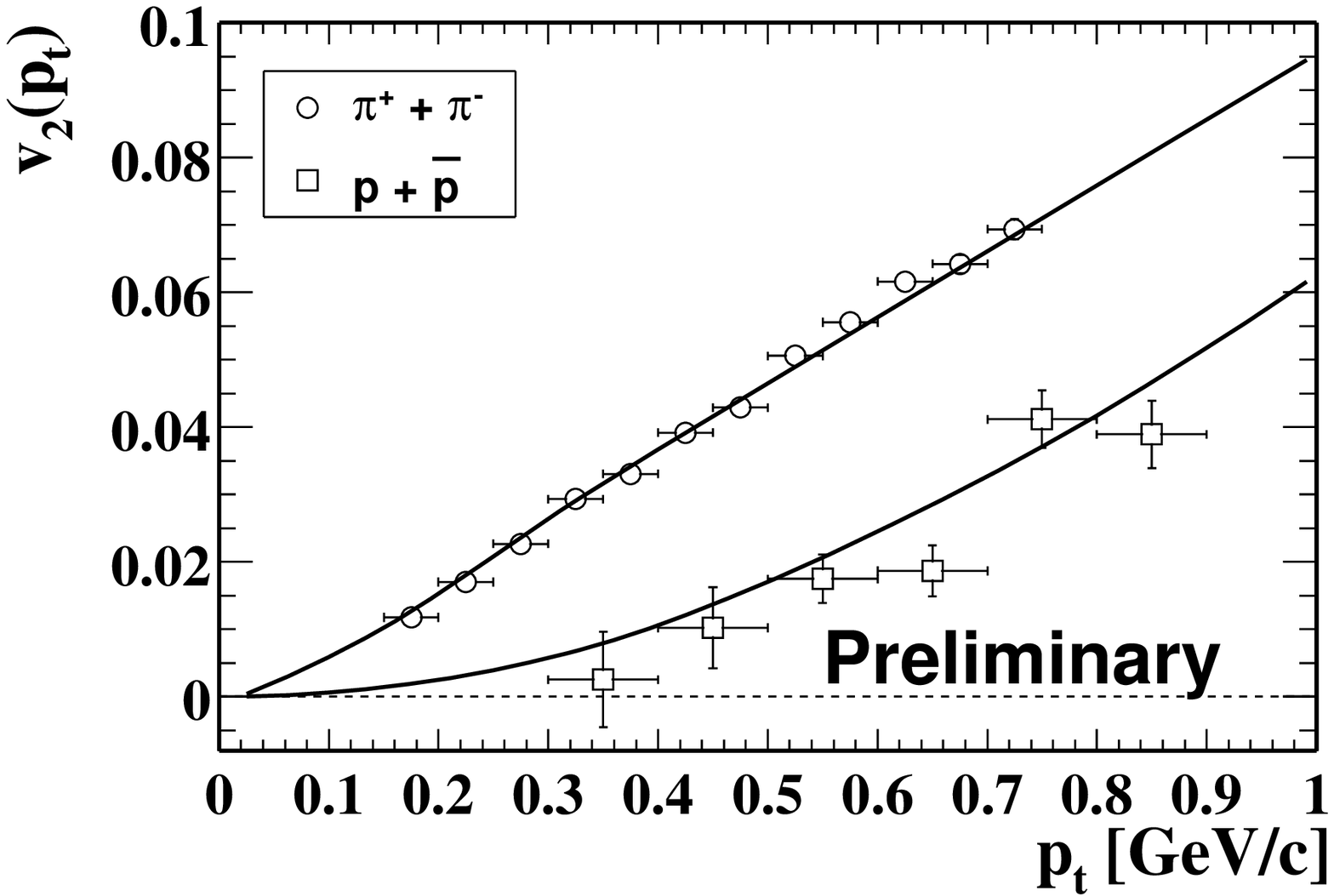}
    \caption{$v_2$($p_t$) for pions and protons + anti-protons,
      the solid lines show the comparison with a full hydrodynamical model 
      calculation~\cite{pasi2}.}
    \label{pions_protons}
  \end{minipage}
  \hspace{\fill}
  \begin{minipage}[t]{0.49\textwidth}
    \includegraphics[width=1.1\textwidth]{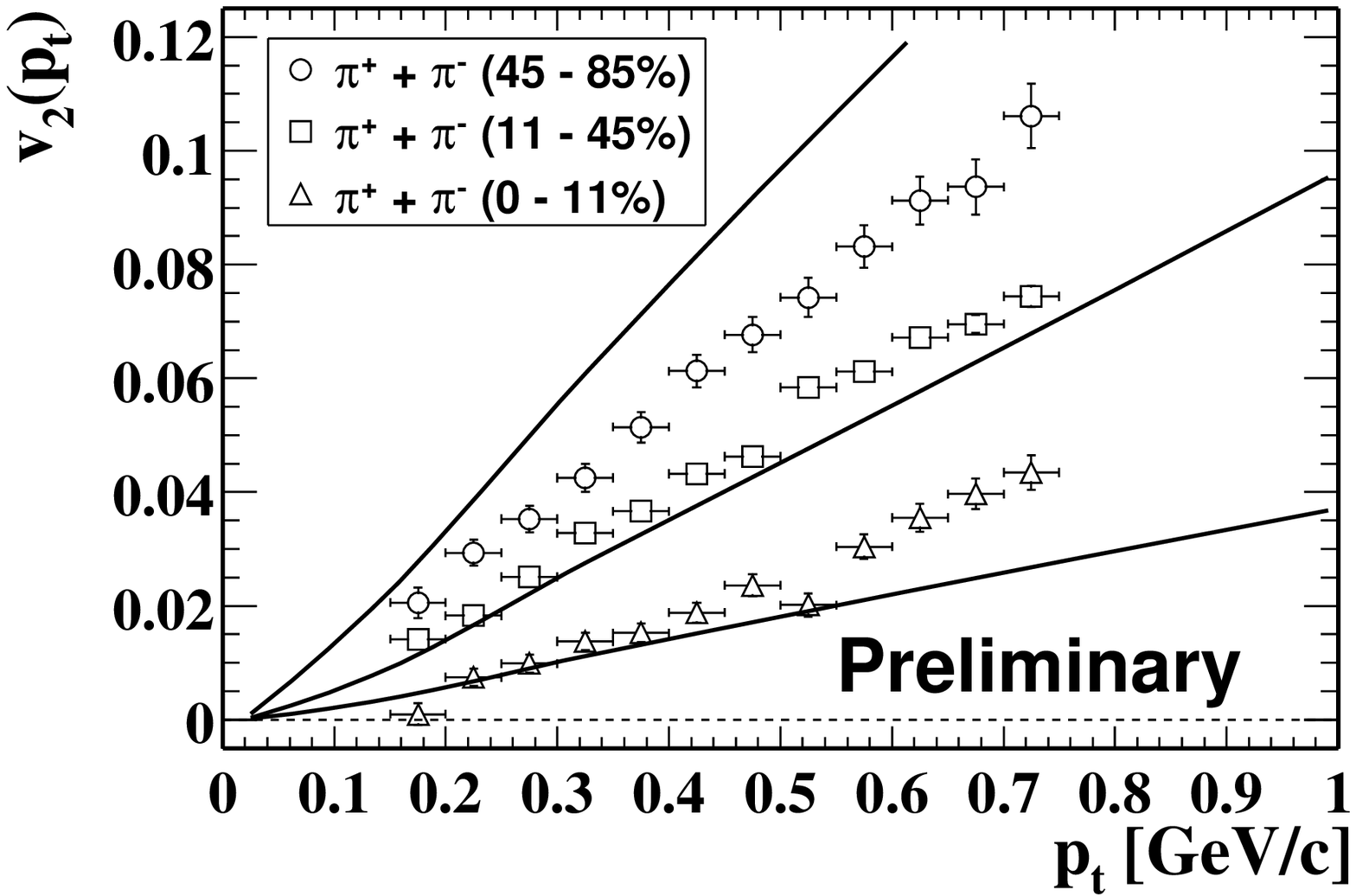}
    \caption{$v_2$($p_t$) for pions in different centralities,
      the solid lines show the comparison with a full hydrodynamical model 
      calculation~\cite{pasi2}.}
    \label{pions_cen}
  \end{minipage}
\end{figure}
\vspace{-0.7cm}

The pions, protons and anti-protons, were selected according to
specific energy loss ($dE/dx$) in the TPC in the momentum range of
0.175 -- 0.75 GeV/$c$, 0.5 -- 0.9 GeV/$c$ and 0.3 -- 0.9 GeV/$c$
respectively. The protons up to a momentum of 0.5 GeV/$c$ were not
used because of proton background from the beam pipe. 
The raw yields of the pions and protons + anti-protons were obtained 
from fitting the $dE/dx$ distributions for
each $y,p_t$ bin with a multiple Gaussian fit and requiring
greater than 90\% purity.

In Fig.~\ref{pions_protons} the differential elliptic flow
as a function of $p_t$ for pions and protons + anti-protons for
minimum-bias collisions is shown.
The uncertainties shown are statistical only.
Using the same assumption to estimate the systematic uncertainties as in
section~\ref{sec:charged} and in addition assuming they are independent
of $p_t$, the systematic uncertainty for minimum-bias data is 13\%.
The positive and negative identified particles, used in this analysis,
have the same $v_2$($p_t$) within statistical uncertainties.
For $v_2$($p_t$) the pions were integrated over $|y| \le 1.0$ and 
the protons + anti-protons over $|y| \le 0.5$.  The pions exhibit an
almost linear dependence of $v_2$($p_t$), whereas the protons +
anti-protons show a more quadratic behavior of $v_2$($p_t$). Such a
behavior is the result of the interplay between the mean expansion
velocity, the anisotropic component of the expansion velocity, and the
thermal velocity of the particles.
A similar effect was
predicted for the case of directed flow~\cite{voloshinv1} at AGS
energies.
The behavior is well described by a full hydrodynamical 
calculation~\cite{pasi2}.
This indicates that the minimum-bias data can be described using
a hydrodynamic description with one set of parameters.

The differential elliptic flow $v_2$ as a function of $p_t$ is plotted
for pions for three different centrality selections in
Fig.~\ref{pions_cen}.
The open triangles, represent the most central
11\% of the measured cross section\footnote{The total measured cross 
section is estimated to correspond to about 80\% to 90\% of the 
geometric cross section, with losses mainly due to vertex finding 
inefficiencies for the low multiplicity events.}. 
The open squares correspond to 11 -- 45\% 
and the open circles correspond to 45 -- 85\% of the measured
cross section. The uncertainties on the points are statistical only. 
The systematic uncertainty is estimated to be 20\% for the most central bin, 
8\% for the mid-central bin and 22\% for the most peripheral bin.
In Fig.~\ref{pions_cen} the lines show the comparison of the
same hydrodynamical calculation for the different centralities.
This comparison shows that the most central and the mid-peripheral 
data can be reasonably well described by this model. 
However, the most peripheral data clearly deviate from the hydrodynamical 
model predictions, indicating a lower degree of thermalization during 
the early stage of the collision for the most peripheral collisions.

\section{Conclusions}

We have made the first measurement of identified particle elliptic
flow at RHIC.
The measured elliptic flow as a function of $p_t$ and centrality 
differ significantly for particles of different masses.
This mass dependence of $v_2$($p_t$) is in close agreement with full 
hydrodynamic model calculations, suggesting that the system for
central and mid-peripheral collisions is close to early
local thermal equilibrium followed by hydrodynamic expansion.
The observed deviation from hydro calculations at $p_t >$ 1.5 GeV/$c$
indicates the transverse momentum region where thermal
equilibrium is not reached. This region can be used to study 
parton energy loss in dense hadronic matter.

\end{document}